\documentclass[12pt]{article}

\let\ssection=\section
\renewcommand{\section}{\setcounter{equation}{0}\ssection}

\def\parag{\hfil\break} 
\def\kikezd{\parag\underbar}
\def\p{{\partial}}

\def\hd{\hat{\delta}}
\def\vb{{\vec b}}

\def\vD{{\vec D}}

\def\vD{{\vec D}}
\def\vx{{\vec x}}

\def\vnabla{{\vec\nabla}}
\def\vA{{\vec A}}
\newcommand\half{{\scriptstyle{\frac{1}{2}}}}
\def\smallcirc{{\raise 0.5pt \hbox{$\scriptstyle\circ$}}}

\begin{document}

\setlength{\baselineskip}{16pt}

\title{Moving vortices in noncommutative gauge theory}

\author{
P.~A.~Horv\'athy\footnote{e-mail: horvathy@univ-tours.fr}\\
Laboratoire de Math\'ematiques et de Physique Th\'eorique\\
Universit\'e de Tours\\
Parc de Grandmont\\
F-37 200 TOURS (France)
\\
and
\\
P.~C.~Stichel\footnote{e-mail: peter@Physik.Uni-Bielefeld.DE}\\
An der Krebskuhle 21\\
D-33 619 BIELEFELD (Germany)
}

\date{\today}

\maketitle

\begin{abstract}
    Exact time-dependent solutions of nonrelativistic
    noncommutative Chern - Simons gauge theory are
    presented in closed analytic form.
    They are different from (indeed orthogonal to) those
    discussed recently by Hadasz, Lindstr\"om, Ro\v cek 
    and von Unge. Unlike theirs, our
    solutions can move with an arbitrary constant velocity, and
    can be obtained from the previously known static solutions
    by the recently found ``exotic'' boost symmetry.
\end{abstract}
\vskip5mm
\noindent\texttt{hep-th/0311157}

\section{Introduction}

In \cite{Had} Hadasz et al. present a time
dependent solution to  
noncommutative Chern-Simons gauge theory in $2+1$ dimensions. 
They introduce the complex operators 
$c={(2\theta)}^{-1/2}(x_{1}-ix_{2})$
and
$K={(2\theta)}^{-1/2}(X_{1}-iX_{2})$ where $X_{i}$ is the covariant
position operator $X_{i}=x_{i}-\theta\epsilon_{ij}A_{j}$ and
$\theta$ is the noncommutative parameter.
Then, generalizing the static one-soliton \cite{LMS},
they posit the Ansatz
\begin{equation}
    K=z(t)\,|0\!><\!0|+S_{1}cS_{1}^\dagger,
    \quad
    \phi=\sqrt{\frac{\kappa}{\theta}}\,|0><\!\varphi(t)|,
    \quad
    A_{0}=-\frac{1}{2\kappa}\phi\phi^\dagger
\label{HadAnsatz} 
\end{equation}
where $\kappa$ denotes the Chern-Simons coupling constant, 
$|\varphi\!>$ some normalized state, and
$z(t)=z_{0}+vt$. Expanding $|\varphi\!>$ into the orthonormal basis
of states $|n_{z}\!>=e^{zc^{\dagger}-\bar{z}c}|n\!>$,
they show that
\begin{equation}
     |\varphi\!>=Me^{i\alpha t}\left(|0_{z}\!>+\sum_{n=1}^\infty
     (-i)^n\frac{d_{n}}{\sqrt{n!}}e^{in\gamma}|n_{z}\!>\right),
     \qquad
     \alpha=\frac{i}{2}\big(\bar{z}_{0}v-z_{0}\bar{v}\big)
     \label{Hadsol}
\end{equation}
where $M$ is a constant
can be a solution, {\it provided} the square of the
velocity takes in\-teger values, $|v|^2=N$.
(\ref{Hadsol}) represents therefore a vortex moving with 
constant quantized speed.

In field theory, a
natural way of producing new solutions from old ones is
by a symmetry transformation.
For example, in commutative Chern-Simons theory,
boosting a static non-relativistic vortex
yields another one which moves with constant speed \cite{JP}.
An infinitesimal boost, $\delta\vx=\vb t$, is implemented
according to
\begin{equation}
    \delta\phi=i(\vb\cdot\vx)\phi-t\vb\cdot\vnabla\phi
    \label{infordboost}
\end{equation}
on the [scalar] matter field, supplemented with the
transformations $\delta A_{\mu}=L_{\delta\vx}A_{\mu}$
of the gauge field. For a finite (\ref{infordboost}) integrates to
\begin{equation}
    \phi^{b}(\vx,t)=e^{i[\vb\cdot\vx-\half\vb^2t]}\phi(\vx-\vb t).
    \label{finordboost}
\end{equation}
The group acts hence through a $1$-cocycle,
\begin{equation}
    U_{\vb}\,\phi(\vx,t)=g(\vb;\vx,t)\phi(\vx-\vb t,t),
    \qquad
    g(\vb;\vx,t)=e^{i[\vb\cdot\vx-\half\vb^2t]}
    \label{ordcocycle}
\end{equation}
which realizes the standard one-parameter central extension of the
Galilei group. The boosts (\ref{infordboost}) commute,
$\left[\delta_{\vb},\delta_{\vb'}\right]=0$.

If $\phi$ is a static vortex, (\ref{finordboost}) clearly 
represents a vortex which moves with velocity $\vb$.

Hadasz et al. \cite{Had} claim that their solution can not
come from boosting the static solution of \cite{LMS} since,
they argue, Galilean invariance is broken in noncommutative
field theory. But the Galilean symmetry can be restored
by a suitable implementation \cite{HMS}; below we
use this latter to construct new, time dependent 
vortex solutions.
    
\section{An exact analytic solution}

For the Ansatz (\ref{HadAnsatz}) 
with $z(t)=z_{0}+vt$, using $\psi$ instead of $\varphi$,
the equations of motion read
\begin{eqnarray}
    -i\theta\dot{\psi}=\big(c^\dagger-\bar{z}(t)\big)
    \big(c-z(t)\big)\psi,\hfill
    \label{eqmot}
    \\[3pt]
    i\theta\dot{z}(t)
    =z(t)-<\!\psi|c|\psi\!>.\hfill
    \label{conscond}
\end{eqnarray}

Now a straightforward calculation shows that
\begin{equation}
    \psi(t)=
    \displaystyle{e^{-\half(\theta^2|v|^2+|z_{0}|^2)
    -i\bar{v}z_{0}\theta}}
    \,e^{-(tv\bar{z}_{0}+\half|v|^2t^2)}
    \,e^{c^\dagger(z(t)-iv\theta)}\big|0\big>
    \label{oursolution}
\end{equation}
provides us with an {\it exact, analytic solution, valid for any complex}
$v$. We find indeed that $\psi(t)$ is an eigenstate of $c$ 
and therefore  of $c-z(t)$ also,
\begin{equation}
    \big(c-z(t)\big)\,e^{c^\dagger(z(t)-iv\theta)}\big|0\big>=
    -iv\theta\,e^{c^\dagger(z(t)-iv\theta)}\big|0\big>.
    \label{hat}
\end{equation}

As the time derivative of $\psi$ is
$\dot{\psi}=v(c^\dagger-\bar{z})\psi$,  (\ref{eqmot})
follows. (\ref{oursolution}) is normalized,
$$
\Big\Vert\psi(t)\Big\Vert^2=
e^{-|z(t)-iv\theta|^2}\Big\Vert
e^{c^\dagger(z(t)-iv\theta)}\big|0\big>\Big\Vert^2=1
$$
upon using the Baker-Campbell-Hausdorff (BCH) formula.
But $c\,\psi=(z(t)-iv\theta)\psi$ by (\ref{hat}),
so that (\ref{conscond}) holds also. Using twice the
BCH formula, (\ref{oursolution}) is conveniently
rewritten also as
\begin{equation}
 \psi(t)=e^{i(\alpha-\theta|v|^2)t}
    \,e^{-i\theta(\bar{v}z_{0}+v\bar{z}_{0})}
    \,e^{-\half\theta^2|v|^2}
    \,e^{(z(t)c^\dagger-\bar{z}(t)c)}
    \,e^{-i\theta vc^\dagger}\big|0\big>.
    \label{secondform}
\end{equation}
\goodbreak

\section{Gauge covariant boosts}

It has been common wisdom that (ordinary) boost invariance
is broken in noncommutative field theory.
Recently we have shown, however, that the Galilean invariance
of NC gauge theory can be restored by implementing the boosts
(infinitesimally) according to
\begin{eqnarray}
    \delta\phi=i\phi\vb\cdot\vx-t\vb\cdot\vnabla\phi,
    \label{phiimp}
    \\[3pt]
    \delta A_{i}=-t\vb\cdot\vnabla A_{i},
    \label{Aimp}
    \\[3pt]
    \delta A_{0}=-\vb\cdot\vA-t\vb\cdot\vnabla A_{0}.
    \label{Animp}
\end{eqnarray}

These formulae look deceivingly similar to (\ref{infordboost});
they are different, though: here all quantities
are {\it operators}, related to functions through the
Weyl correspondence. The operator product understood here
corresponds to the Moyal ``star'' product of functions.
The first term in (\ref{phiimp}) corresponds, e. g., to right 
star-multiplication, $\phi\star(i\vb\cdot\vx)$, cf. \cite{HMS}.

Owing to the operator commutation relation
$\big[x_{1},x_{2}\big]=-i\theta$, the boosts do not commute,
but satisfy rather the ``exotic'' commutation relation
of the two-fold centrally extended planar Galilei group
\cite{HMS,LL},
\begin{equation}
    \big[\delta_{\vb},\delta_{\vb'}\big]\phi
    =-i\theta\,(\vb\times\vb')\,\phi,
    \label{exocommrel}
\end{equation}
as follows at once from (\ref{phiimp}).
Finite boosts are represented according to
\begin{equation}
    U_{\vb}\,\phi(\vx,t)=\phi(\vx-\vb t,t)g(\vb;\vx,t),
    \qquad
    g(\vb;\vx,t)=e^{i(\vb\cdot\vx-\half\vb^2 t)}
    \label{NCcocycle}
\end{equation}
which looks again similar to (\ref{ordcocycle}),
but the cocycle $g(\vb;\vx,t)$ here is {\it operator 
valued}, and acts from the right by operator multiplication.

Note that in terms of functions (\ref{NCcocycle}) would mean,
by the Weyl correspondence, 
\hfill\break
$\phi(\vx-\vb t,t)\star 
(\exp_{\star}\left[i(\vb\cdot\vx-\half\vb^2t)\right])$,
where $\exp_{\star}$ is the exponential w. r. t. the star
product.

With this implementation, the boosts act as symmetries~:
they carry any solution of the equations of motion into
some (other) solution. But can either of the
solutions be obtained in this way ?  At first sight, the
answer seems to be negative: a boost changes $A_{0}$,
while our Ansatz requires it to be fixed. This can easily
be cured, though, if we supplement it with a
suitable gauge transformation, namely by $\Lambda=t\vb\cdot\vA$.
This amounts to replacing 
(\ref{phiimp})-(\ref{Aimp})-(\ref{Animp})
by {\it gauge-covariant expressions} \cite{JP2},
i. e. by
\begin{eqnarray}
 \hd\phi=i\phi\vb\cdot\vx-t\vb\cdot\vD\phi,
    \label{gcphiimp}
    \\[3pt]
    \hd A_{i}=-t\vb_{k}\epsilon_{ki}B=
    -\frac{t}{\kappa}b_{k}\epsilon_{ki}\,\phi\phi^\dagger,
    \label{gcAimp}
    \\[3pt]
    \hd A_{0}=-tb_{k}F_{k0}=
    -\frac{t}{\kappa}b_{k}\epsilon_{ki}J_{i},
    \label{gcAnimp}
\end{eqnarray}
where $D_{i}=\p_{i}-iA_{i}$ and the field-current identities 
$\kappa B=-\phi\phi^\dagger$ and $\kappa E_{i}=\epsilon_{ik}J_{k}$
(which belong to the CS+ matter field equations 
\cite{LMS, HMS}) were also used. 
Eq. (\ref{gcAimp}) implies that
$
\hd K=\frac{\theta t}{\kappa}\,b\,\phi\phi^\dagger.
$

How does the Ansatz (\ref{HadAnsatz}), behave with respect
to our gauge-covariant boosts~?
Using $K\phi=z(t)\phi$ and $K^\dagger\phi=\bar{z}(t)\phi$, we find,
putting $b=(2\theta)^{-1/2}\big(b_{1}-ib_{2})$,
\begin{eqnarray}
    \hd\phi=i\theta\phi(bc^\dagger+\bar{b}c)+
    \phi\,t\big\{\bar{b}(c-z(t))-b(c^\dagger-\bar{z}(t))\big\}
\end{eqnarray}
i. e.
\vskip-5mm
\begin{eqnarray}
    \hd\psi=
    \left[-i\theta(\bar{b}c+bc^\dagger)
    +t\big\{(b(c^\dagger-\bar{z}(t))-\bar{b}(c-{z}(t))\big\}
    \right]\psi(t)
    \label{infpsiboost}
\end{eqnarray}
completed with
$ \hd K=tb|0\!><\!0|$. 
Hence
$$ 
K+\hd K=\big(z_{0}+(v+b)t\big)|0\!><\!0|+S_{1}cS_{1}^\dagger
$$ 
i. e., the boost parameter $b$ merely adds to the velocity,
as expected, while
$A_{0}$ is indeed invariant $\hd A_{0}=0$. 
Finally, by (\ref{hat}),
\begin{equation}
    \hd\psi=
    \left[-i\theta\big\{\bar{b}(z(t)-iv\theta)+bc^\dagger\big\}
    +t\big\{
    b(c^\dagger-\bar{z}(t))+iv\theta\bar{b}\big\}\right]\psi(t).
    \label{311}
\end{equation}
Putting $v\to v+b$ into
(\ref{oursolution}) and expanding to lowest order in $b$
we obtain instead 
\begin{equation}
    \psi_{v+b}\simeq\psi_{v}+\hd\psi_{v}
    +\half\big(\bar{b}v-b\bar{v}\big)(\theta^2-t^2)\psi_{v}.
    \label{infimp}
\end{equation}

To explain this result, let us remind that,
in gauge-covariant framework,
the commutators close up to 
 gauge transformations \cite{JP2} only. 
Using the Gauss constraint 
$\kappa B=-\phi\phi^\dagger$ and the Ansatz (\ref{HadAnsatz}),
instead of (\ref{exocommrel}) we find indeed
\begin{equation}
    \big[\delta_{b},\delta_{b'}\big]\phi
    =\,\big(b\bar{b'}-\bar{b}b'\big)(\theta^2-t^2)\phi.
     \label{gccomrel}
\end{equation}

The first factor here is the vector product, already 
present in the ``exotic'' relation (\ref{exocommrel}).
The
second factor is a remnant of the 
follow-up gauge transformation.
From (\ref{gcphiimp}),
the follow-up gauge transformation is indeed
given, for a generic $\phi$,
by $\theta t^2(\bar{b}v-b\bar{v})B\phi$.
Then using the Gauss constraint and (\ref{HadAnsatz}),
the second term in (\ref{gccomrel}) follows.

Let us now turn to finite boosts. Exponentiating (\ref{infpsiboost})
we get
\begin{equation}  
 \hat{U}_{b}\psi_{v}
 =e^{-i\theta(\bar{b}c+bc^\dagger-|b|^2t)}
 e^{t(b(c^\dagger-\bar{z}(t))-\bar{b}(c-z(t))}\psi_{v},
\label{fimplementation}
\end{equation}
where the additional term in the first exponential
was introduced by analogy with the ordinary case (\ref{finordboost}).
Using the BCH formula and (\ref{hat}),
(\ref{fimplementation})
can also be presented as
$$
\psi^{b}_{v}=e^{-t(b\bar{z}(t)-\bar{b}z(t))}
\,e^{-(z(t)-iv\theta)\bar{b}(t+i\theta)}
\,e^{-\half|b|^2(\theta^2+t^2)}
\,e^{c^\dagger b(t-i\theta)}
\psi_{v}.
$$
We conclude that a finite boost acts on our solution
(\ref{oursolution}) according to 
\begin{equation}    
    \hat{U}_{b}\psi_{v}
    =\psi_{v+b}\,
    e^{-\half(\bar{b}v-b\bar{v})(\theta^2-t^2)}.
    \label{fboost}
\end{equation}

This provides us with the composition law for 
gauge-covariant boosts as acting on
our moving vortex (\ref{oursolution}).

Chosing $v=0$ in (\ref{fboost})
yields furthermore that the moving solutions (\ref{oursolution})
can be obtained from the static one, $\psi_{0}$ of \cite{LMS},
by a gauge-covariant boost, 
\begin{equation}
    \psi_{v}=\hat{U}_{v}\psi_{0},
\end{equation}
as expected. 
By (\ref{fboost}), moving vortices are, however, only 
up-to-phase invariant under further boost.
The additional phase is understood by putting $\hat{U}_{b}=e^{\hd_{b}}$. 
Then the BCH formula yields 
$
\displaystyle{
e^{\hd_{b}}e^{\hd_{v}}=e^{\hd_{b}+\hd_{v}+\half[\hd_{b},\hd_{v}]}
}
$
which, together with (\ref{gccomrel}), allows us to infer
$$
    \hat{U}_{b}\psi_{v}=\hat{U}_{b}\hat{U}_{v}\psi_{0}
    =
    \hat{U}_{b+v}e^{-\half(\bar{b}v-b\bar{v})(\theta^2-t^2)}
    \psi_{0}
$$ 
which is (\ref{fboost}).
(\ref{infpsiboost}) and (\ref{311})  imply in fact that
$
\hd_{b+v}=\hd_{b}+\hd_{v}
$.

 The infinitesimal version is just (\ref{infimp}).

\section{Comparision of the two types of solutions}

The  solution (\ref{Hadsol}) of Hadasz et al. \cite{Had} 
depends on time through 
the factor $e^{i\alpha t}$, chosen so that
$\dot{\varphi}$ is orthogonal to $\varphi$. 
This yields some involved
recursion relations for the coefficients $d_{n}$
in (\ref{Hadsol}) that only allows 
 $|v|^2=s^2=N$, an integer \cite{Had}.
Our solution has instead
$
e^{i(\alpha t-\theta|v|^2t)}
$
cf. (\ref{secondform}).
To derive the consequences, let us
put $v=se^{i\gamma}$ $s\geq0$, and 
expand into basis states, using
$$
e^{-i\theta vc^\dagger}|0\!>
=
\sum_{n=0}^\infty(-i)^n\frac{(\theta v)^n}
    {\sqrt{n!}}\,|n\!>.
$$
This allows us to 
present our solution in the form
\begin{equation}
    \psi(t)=e^{i(\alpha t-t\theta|v|^2)}
    e^{-i\theta(\bar{v}z_{0}+v\bar{z}_{0})}
    e^{-\half\theta^2|v|^2}
    \sum_{n=0}^\infty(-i)^n\frac{(\theta s)^n}
    {\sqrt{n!}}e^{in\gamma}\,|n_{z}\!>.
    \label{thirdform}
\end{equation}
 Our expansion coefficients $(\theta s)^n$ obey a simple
recursion relation, 
allowing our solution to have any velocity $v$.

What is the relation of the two types of solutions
associated with the same quantized value $|v|^2=N$ of the velocity~?
Due to the orthogonality of the $|n_{z}\!>$ states,
 putting $d_{0}=1$
and $\theta=1$, we find for their scalar product
$$
\big<\psi\,\big|\,\varphi\big>
\propto
\sum_{n=0}^\infty\frac{d_{n}}{n!}(\sqrt{N})^n,
$$
which is just the value at $\zeta=\sqrt{N}$
of the generating function determined by Hadasz et al.,
$$
G(\zeta)=\sum_{n=0}^\infty\frac{d_{n}}{n!}\zeta^n
=
e^{\sqrt{N}\zeta}\left(1-\frac{\zeta}{\sqrt{N}}\right)^N
$$
cf. \# (3.29), (3.31) of Ref. \cite{Had}.
But $G(\sqrt{N})=0$, so 
the two states are indeed orthogonal to each other~!

\section{Conclusion}

We have found 
an exact time dependent analytic solution of noncommutative 
gauge theory. It represents a $1$-soliton
travelling with arbitrary constant speed, and can be obtained
from the static $1$-soliton by a gauge-covariant 
``exotic'' boost. 
While finite gauge-covariant transformations
are not in general known in closed form \cite{JP2},
in our particular case we did find such an expression.
  
For the particular quantized values 
$|v|^2=N$ of the velocity, 
when both types of solutions exist, ours are orthogonal
to those of Hadasz et al. \cite{Had}.

\kikezd{Acknowledgments}
We are indebted to L. Martina for his interest. 
After our paper was completed and submitted,
Hadasz et al.
have submitted a revised version of their
paper, where they also find, independently,
the boosted solutions discussed here.
We would like to thank also
the Authors of \cite{Had}  for correspondence.


\end{document}